\documentclass[10pt]{article}
\usepackage[OE]{express}
\usepackage{subfig}
\usepackage{hyperref}

\begin{document}

\title{Application of optical single-sideband laser in Raman atom interferometry}


\author{Lingxiao Zhu,\authormark{1} Yu-Hung Lien,\authormark{1,*} Andrew Hinton,\authormark{1,2} Alexander Niggebaum,\authormark{1} Clemens Rammeloo,\authormark{1} Kai Bongs,\authormark{1} and Michael Holynski\authormark{1,3}}

\address{\authormark{1}School of Physics and Astronomy, University of Birmingham, Edgbaston, Birmingham B15 2TT, UK \\
\authormark{2}Quantum Metrology Laboratory, RIKEN, Wako, Saitama 351-0198, Japan \\
\authormark{3}m.holynski@bham.ac.uk}
\email{\authormark{*}y.lien@bham.ac.uk}



\begin{abstract}
A frequency doubled I/Q modulator based optical single-sideband (OSSB) laser system is demonstrated for atomic physics research, specifically for atom interferometry where the presence of additional sidebands causes parasitic transitions. The performance of the OSSB technique and the spectrum after second harmonic generation are measured and analyzed. The additional sidebands are removed with better than 20~dB suppression, and the influence of parasitic transitions upon stimulated Raman transitions at varying spatial positions is shown to be removed beneath experimental noise. This technique will facilitate the development of compact atom interferometry based sensors with improved accuracy and reduced complexity.
\end{abstract}



\section{Introduction}
\begin{figure}[htp]
	\centering
    \includegraphics[width=0.8\linewidth]{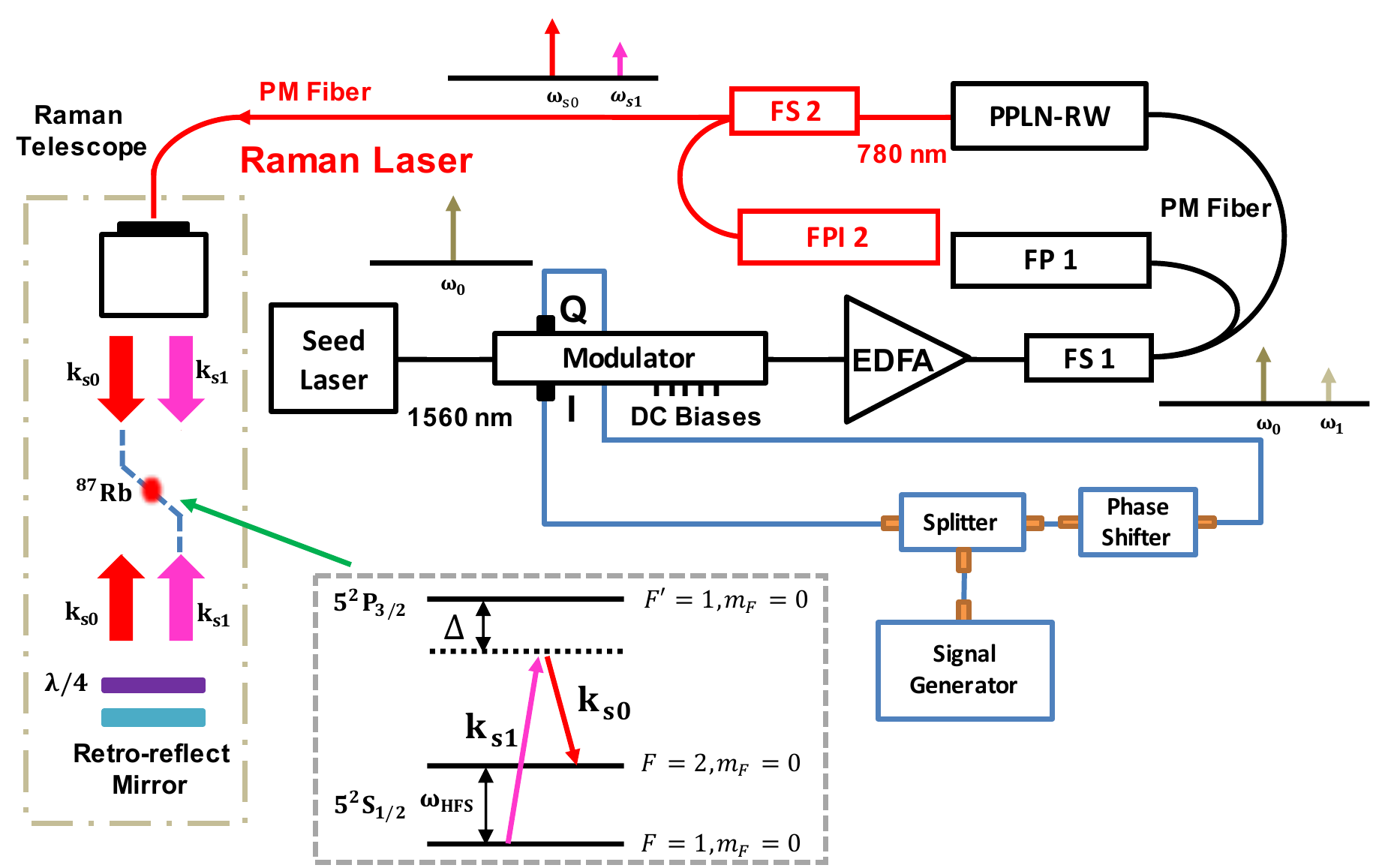}
    \caption{\label{fig:apparatus} Simplified diagram of the apparatus. A simplified \(\mathrm{^{87}Rb}\) level diagram is included as an inset. The pair of the counter-propagating beams \(\textbf{k}_{s0} \downarrow\) and \(\textbf{k}_{s1} \uparrow\) are resonant with a two-photon Raman transition between the magnetically insensitive hyperfine sublevels \(m_{F} = 0\). FS 1: 1560~nm 1:99 fibre splitter. FS 2: 780~nm 1:20 fibre splitter. FPI 1, 2: 1560~nm and 780~nm Fabry-P\'erot Interferometer, respectively. All fibers are polarization-maintaining.}
\end{figure}

Raman atom interferometry (RAI), first demonstrated over 25 years ago~\cite{Kasevich1991}, has enabled the creation of the most precise laboratory measurement devices for absolute gravity~\cite{Freier2016}, gravity gradients~\cite{Sorrentino2010} and rotation~\cite{Gustavson2000}. This makes atom interferometry a promising quantum technology for applications ranging from fundamental physics in space~\cite{Altschul2015, Bongs2015} to practical concerns such as civil engineering~\cite{Hinton2017}. These applications require the translation of laboratory systems into small, robust and low-power devices, which poses particular challenges for the light source. Raman atom interferometers as illustrated in Fig.~\ref{fig:apparatus} use a two-photon transition to couple two hyperfine levels of the atomic ground state. This allows manipulation of the atom to split, redirect and recombine its quantum mechanical wave function, encoding the gravitational phase shift between the two trajectories in the relative atomic populations of the two hyperfine levels. This demands two laser beams with a frequency difference matching the ground state hyperfine splitting, typically several GHz, and the relative optical phase coherence at the mrad level over several hundred ms. 

Laboratory versions of the Raman atom interferometer light source have been implemented through optical phase locking~\cite{LeGouet2009}, acousto-optic modulators~(AOM)~\cite{Bouyer1996} and electro-optic modulators~(EOM)~\cite{Kasevich1991a}. Some state-of-the-art Raman laser systems have been developed for very stringent payload limitation and harsh environment~\cite{Menoret2011, Schkolnik2016}. Recent quantum technology developments focus on robust telecoms technology with electro-optic modulation and frequency doubling~\cite{Cheinet2006, Merlet2014, Theron2015, Muquans}, which in principle allow the realisation of an entire atom interferometer from a single laser source. However, the EOM scheme usually produces a double sideband (DSB) spectrum which contains redundant sidebands. These redundant sidebands not only waste optical power but also introduce extra undesirable interactions which affect the system systematics and impair performance~\cite{Carraz2012}. Here we demonstrate a laser system for RAI, which maintains the compactness and simplicity of EOM systems without the adverse effects of the redundant sidebands.

The system relies on I/Q modulator based optical single-sideband generation (OSSB)~\cite{Buhrer1962, Page1970, Izutsu1981, Cusack2004, Hangauer2014}, with the novel feature of maintaining the single sideband modulation through an optical frequency doubling element. Our OSSB system is shown schematically in Fig.~\ref{fig:apparatus}. The first stage comprises a 1560~nm fibre laser and an I/Q modulator for single sideband generation. The output is then converted into 780~nm by second harmonic generation (SHG) to resonate with the \(\mathrm{^{87}Rb}\) D2~line. The I/Q modulator allows implementing OSSB with a high degree of versatility allowing the optimization of undesired sideband suppression after the optical frequency doubler. The SHG unavoidably pairs up different spectral components in 1560~nm OSSB and converts them into 780~nm, which requires optimization in order to avoid suppressed sidebands in the fundamental band reemerging in the frequency doubled spectrum. We measure the spectra of our OSSB, including both fundamental and SHG. We finally implement the full-carrier optical single sideband scheme (FC-OSSB) as an interrogation laser for RAI. We demonstrate that the FC-OSSB Raman laser suppresses the influence of undesired sidebands, by measuring the spatial dependence of the Rabi frequency and interferometric phase shift and comparing this to that of an EOM modulation scheme. At the end, we include the theory for the I/Q modulator based OSSB, the SHG spectrum of OSSB, and relevant notations in the appendix.

\section{Apparatus}

Our Raman atom interferometer including an FC-OSSB based Raman laser is illustrated in Fig.~\ref{fig:apparatus}. The Raman laser comprises a highly stable 1560~nm fibre laser (Rock\texttrademark, NP Photonics), a fibre-coupled I/Q modulator (MXIQ-LN-40, iXBlue), an erbium-doped fibre amplifier (EDFA) (YEDFA-PM, Orion Laser), and a nonlinear wavelength conversion module (WH-0780-000-F-B-C, NTT Electronics). The I/Q modulator is essentially a dual parallel Mach-Zehnder modulator (MZM)~\cite{Izutsu1981, Shimotsu2001}. The in-phase and quadrature ports of the I/Q modulator are driven by the same RF signal \(\omega_{m}\) except an RF phase shifter is installed on the quadrature port to generate the necessary \(\pi/2\) phase shift. The optical phase shifters \(\Phi_{1,2,3}\) are separately controlled by separate precision voltage sources. The output of the fibre laser is modulated by the I/Q modulator to generate FC-OSSB, which is amplified by the EDFA to 1~W. The nonlinear wavelength conversion module based on a periodically poled lithium niobate ridge waveguide (PPLN-RW) is used to convert the FC-OSSB signal to 780~nm. The optical power of 780~nm light is approximately 440~mW. The spectra of both 1560~nm and 780~nm are simultaneously monitored by two Fabry-P\'erot interferometers (FPI) (SA210-12B, SA200-5B, Thorlabs) with free spectral range (FSR) 10~GHz and 1.5~GHz respectively. The two major frequencies in the 780~nm FC-OSSB spectrum drive a two-photon Raman transition between the \(\mathrm{^{87}Rb}\) ground state hyperfine levels with a detuning \(\mathrm{\Delta = 2\ GHz}\) which is illustrated in the inset of Fig.~\ref{fig:apparatus}. As shown in next section, other sidebands are strongly suppressed compared with these two frequencies.

In the beginning of the experiment, \(\mathrm{^{87}Rb}\) atoms are loaded into a 3D magneto-optical trap and then launched into the interrogation region by the moving molasses technique~\cite{Hall1989, Lett1989}. The Raman laser beam is directed into the interrogation region and retro-reflected by a mirror. The light field inside the interrogation region consists of two counter-propagating beams which contain two frequencies. In principle, there are four different combinations that can drive the two-photon Raman transition used in RAI. Nevertheless, only one pair, as illustrated in Fig.~\ref{fig:apparatus}, is selected by Doppler selection and the polarization. When the atoms are in the interrogation region a \(\pi/2-\pi-\pi/2\) Raman pulse sequence is applied to perform RAI. The phase shift \(\Delta \Phi_{RAI}\) of RAI is determined by measuring the population ratio of the ground state hyperfine levels. In the case of a local gravity acceleration \(g\), \(\Delta \Phi_{RAI}\) is given by:

\begin{equation}\label{eq:phase shift}
	\Delta \Phi_{RAI} = \cos^{-1}(\frac{P_{1}}{P_{1} + P_{2}}) = (\textbf{k}_{\mathrm{eff} }\cdot \textbf{g} - \alpha)T^{2}
\end{equation}
where \(P_{1,2}\) are the populations of the ground state hyperfine levels F = 1 and 2, the interferometric time \(T\) is the interval between adjacent Raman pulses, \(\alpha\) is the chirp rate of \(\omega_{m}\) for compensating the Doppler shift, and \(\textbf{k}_{\mathrm{eff}} = \textbf{k}_{s1}\uparrow - \textbf{k}_{s0}\downarrow\) is the effective wave vector~\cite{Kasevich1991}. 

\section{Results}
\subsection{Spectra of OSSB}
The system was then used to demonstrate that FC-OSSB can be realized both before and after the SHG. According to Eq.~(\ref{eq:FC-OSSB}) and (\ref{equ:FC-OSSB-780}), the OSSB spectrum at 780~nm is degraded due to the SHG. The sideband $\omega_{s-1}$ is not eliminated in 780~nm, and is generated by the sum frequency generation (SFG) between the sidebands $\omega_{-2}$ and $\omega_{1}$ in 1560~nm. This is expressed as:

\begin{equation*}
8(1+e^{i\pi/2})J_{2}(\beta)J_{1}(\beta)e^{i(\omega_{-2}+\omega_{1})t}
\end{equation*}

The sideband $\omega_{s2}$ is proportional to the term:

\begin{equation*}
8e^{i\pi/2}[J_{0}(\beta)J_{2}(\beta)e^{i(\omega_{0}+\omega_{2})t}-J_{1}(\beta)J_{1}(\beta)e^{i(\omega_{1}+\omega_{1})t}]
\end{equation*}

\begin{figure}[htp]
	\centering
 	\includegraphics[width=0.7\linewidth]{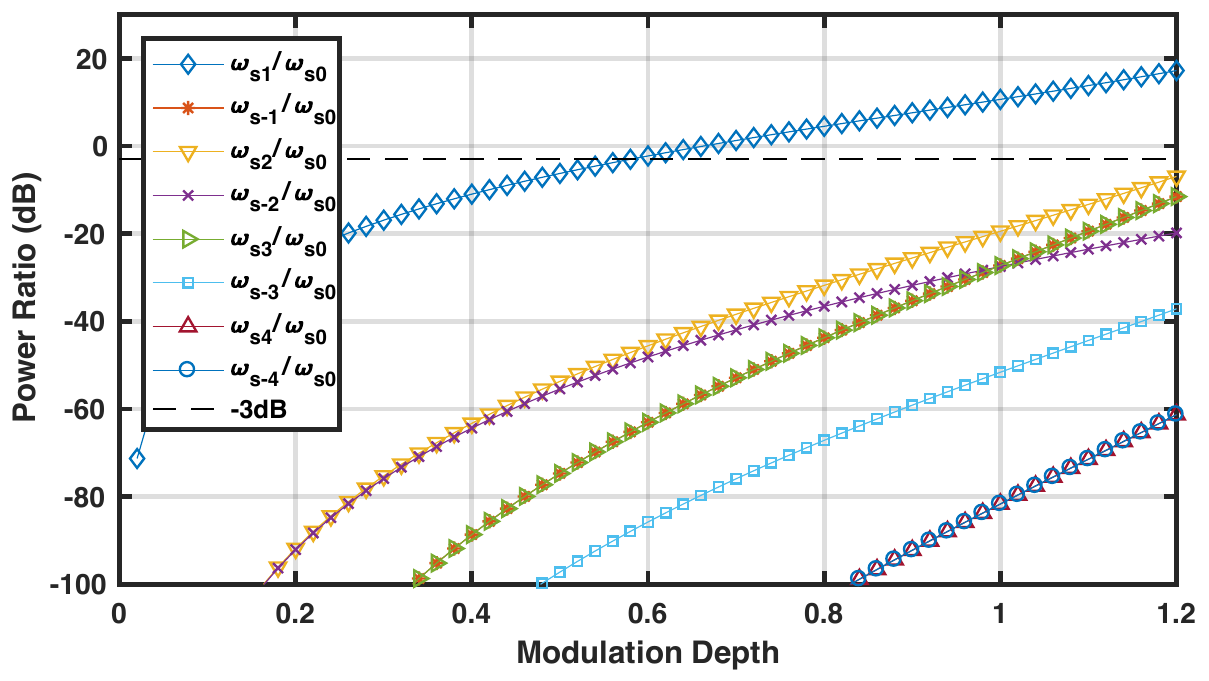}
    \caption{\label{fig:SidebandsSupress} Simulation of the optical power ratio of the sidebands with respect to the carrier after the SHG. The power ratio where is -3~dB is denoted in dashed line.}
\end{figure}

As shown in Fig.~\ref{fig:SidebandsSupress}, the amplitude of these unwanted sidebands depends on the setting of the modulator and the issue is further addressed in the appendix. To optimize the 780~nm FC-OSSB,  We start from its counterpart at 1560~nm by adjusting \(\Phi_{3}\) and observing the suppression pattern of the different frequency components such as \(\omega_{-1}\) and \(\omega_{0}\). Further optimization is achieved by adjusting \(\Phi_{1,2}\). Subsequently, the power and the spectrum of 780~nm OSSB are adjusted primarily through the modulation depth \(\beta\) but also \(\Phi_{1,2}\). Eventually the temperature of the wavelength conversion module is adjusted to shift the gain profile and finish the optimization.

\begin{figure}[htp]
	\centering
	\subfloat[]{\includegraphics[width=0.45\textwidth]{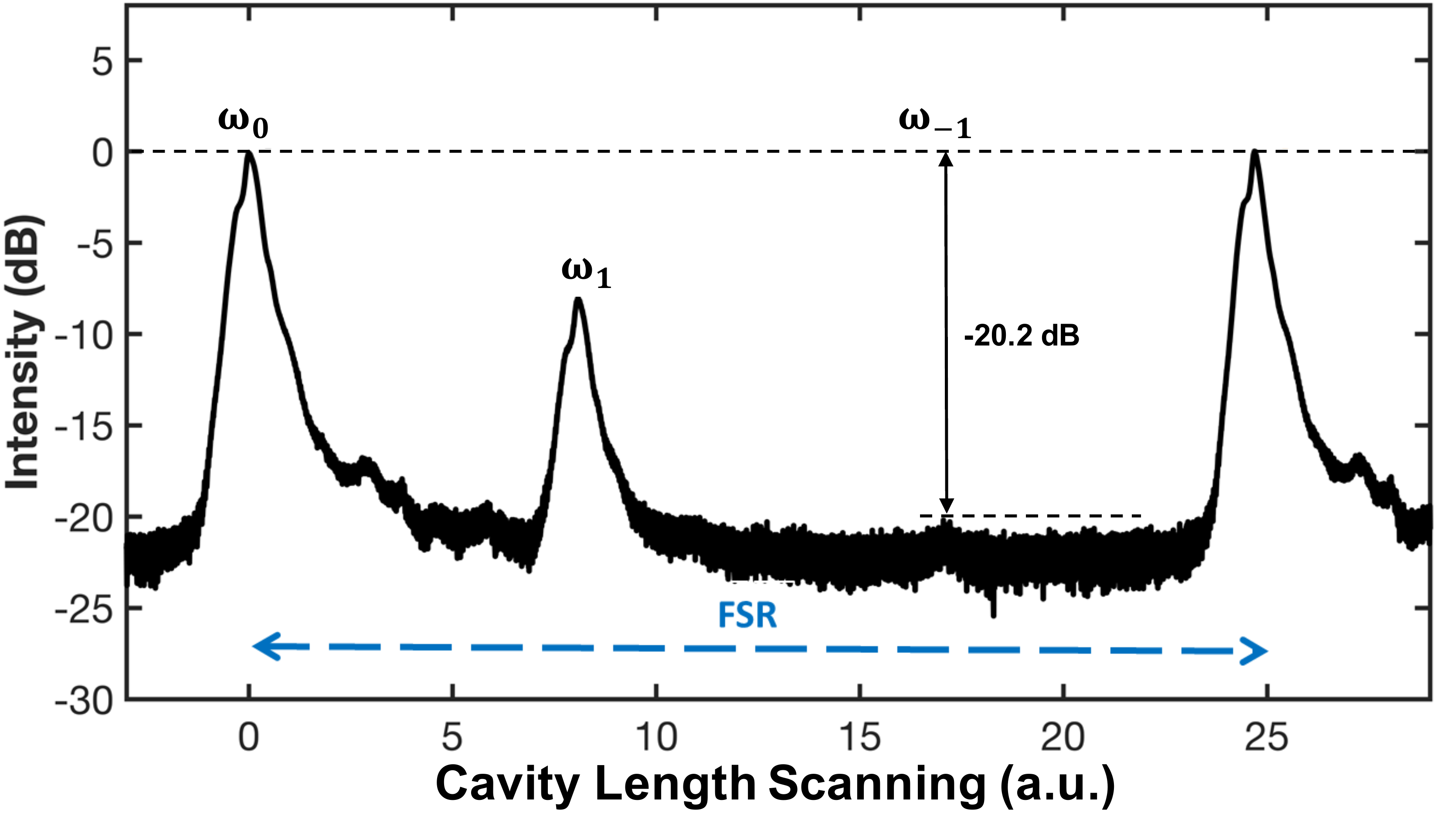}}
	\subfloat[]{\includegraphics[width=0.45\textwidth]{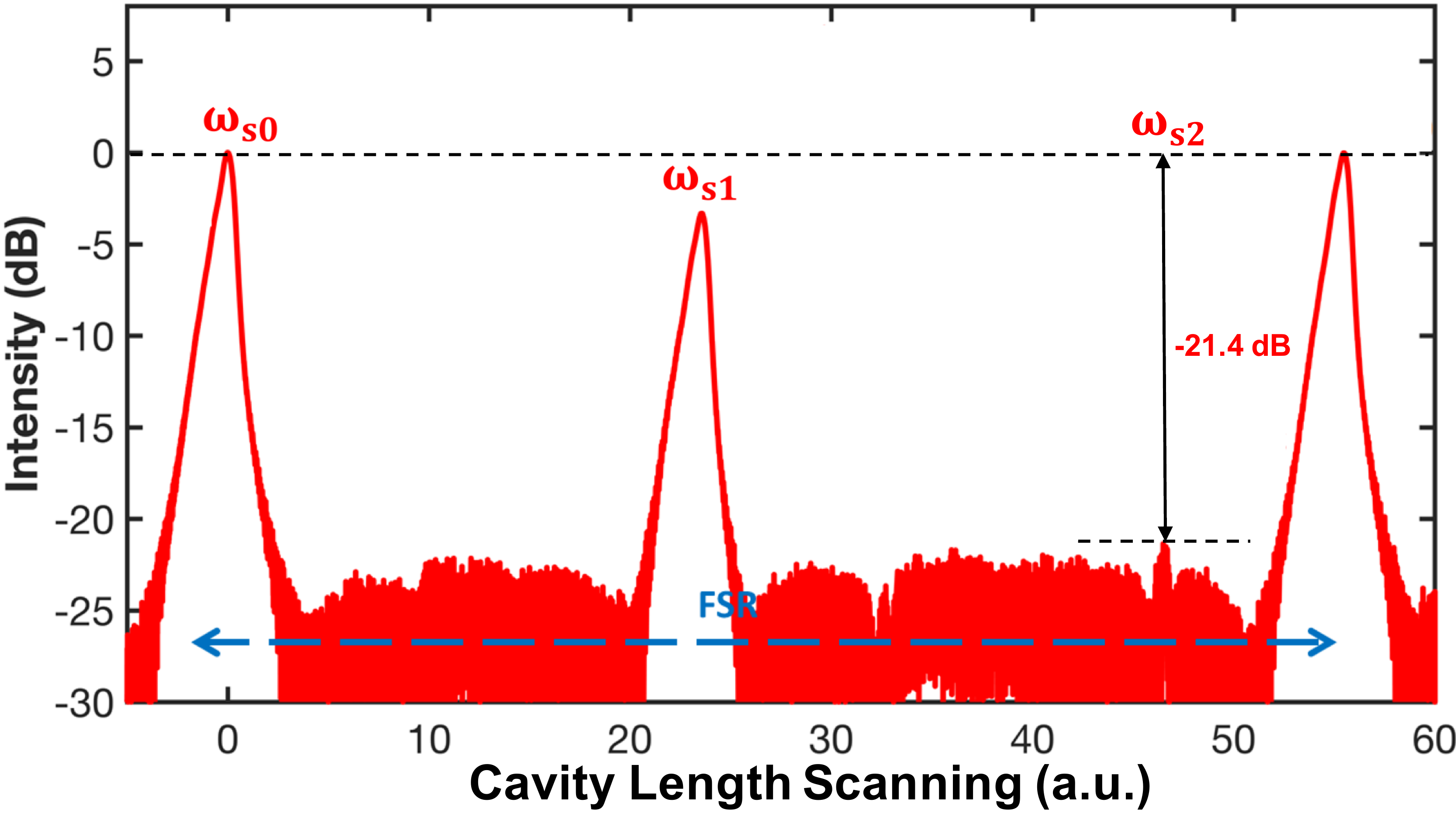}}
	\\
	\subfloat[]{\includegraphics[width=0.45\textwidth]{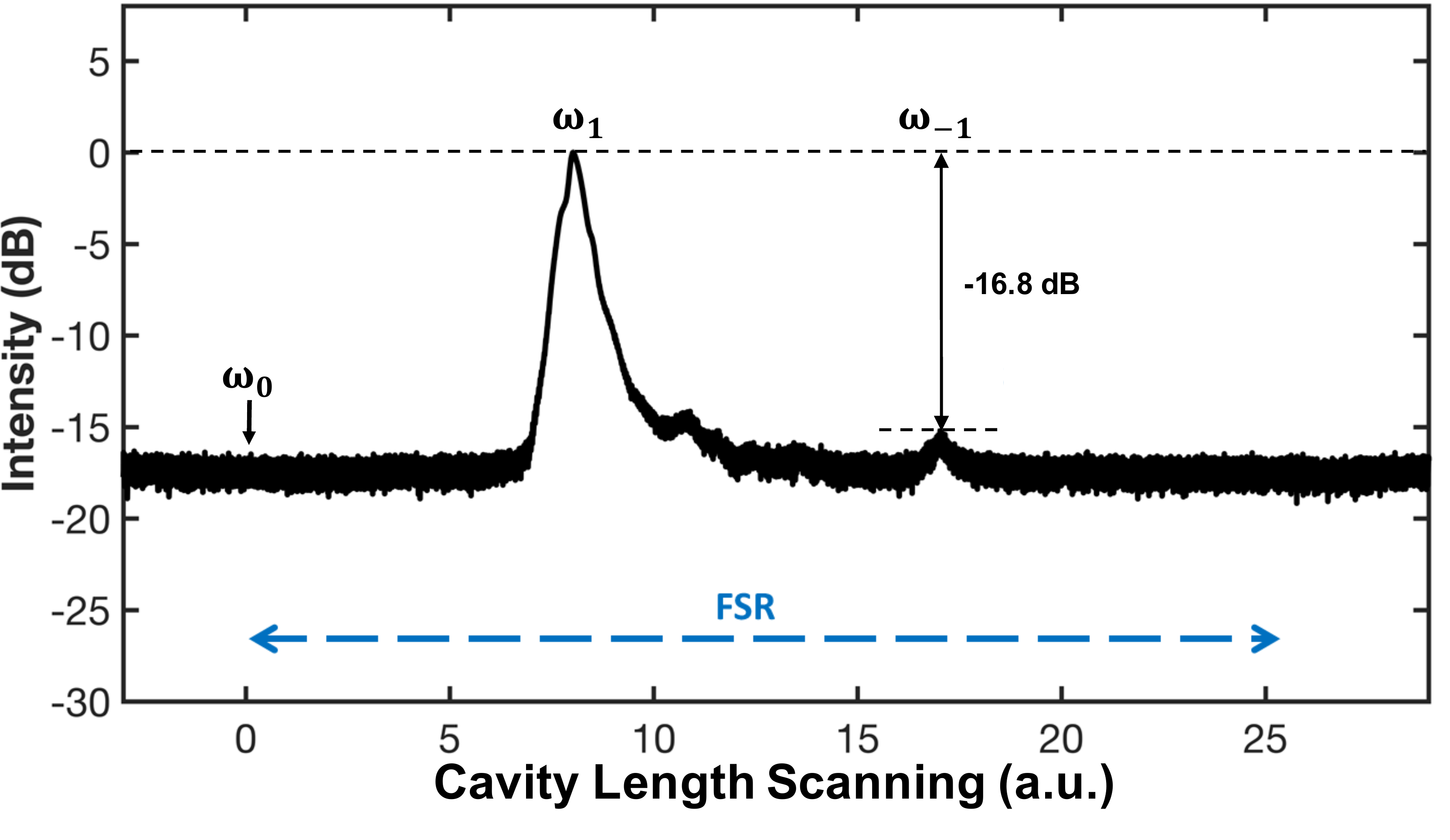}}
	\subfloat[]{\includegraphics[width=0.45\textwidth]{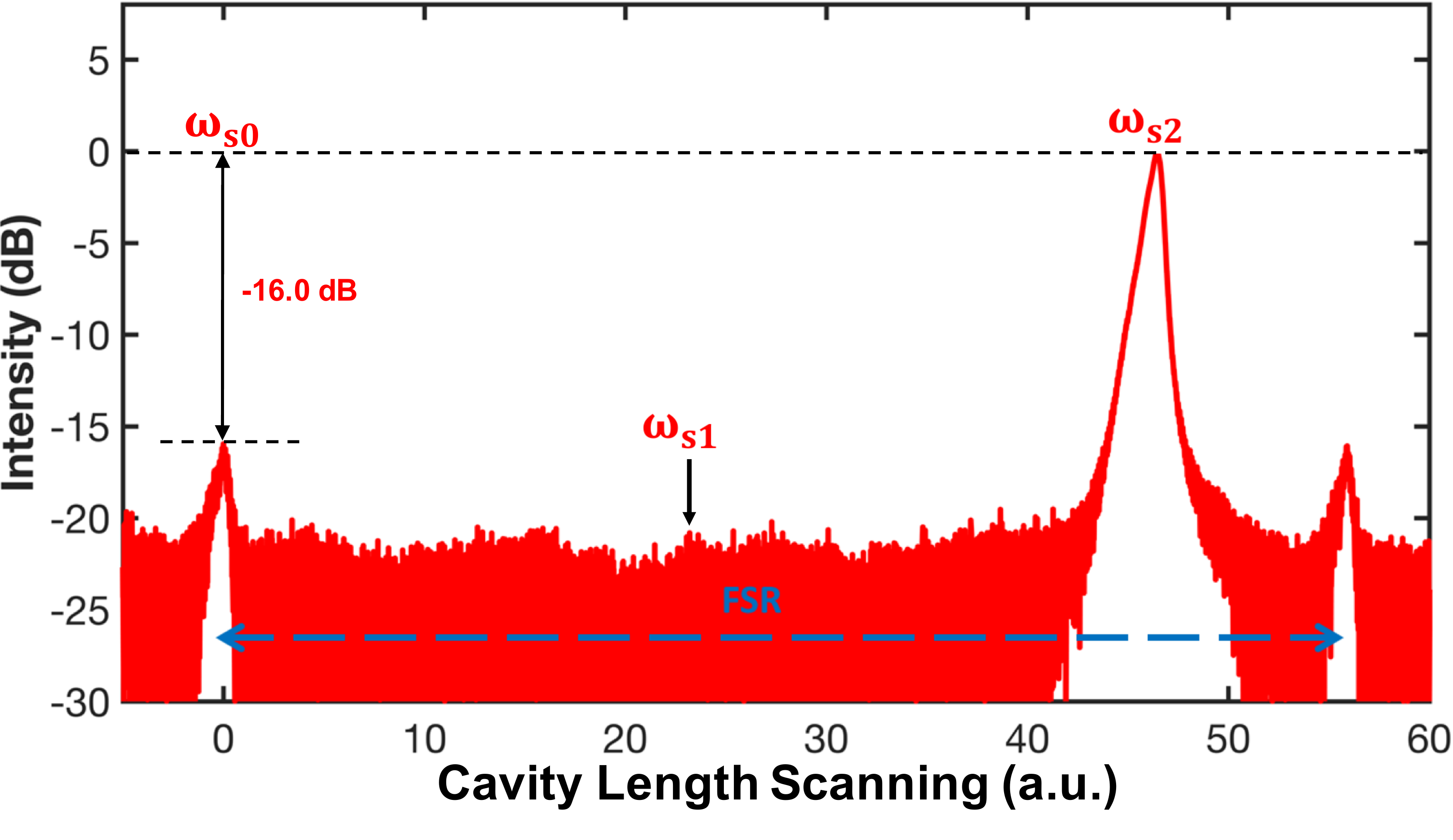}}
	\caption{\label{fig:spctrum_OSSB} The FC- and SC-OSSB spectra. (a) and (c) are the 1560~nm spectra of FC- and SC-OSSB, respectively. (b) and (d) are the 780~nm spectra of FC- and SC-OSSB, respectively. The FSRs of FPIs are marked by the blue dash lines.}
\end{figure}

Figure~\ref{fig:spctrum_OSSB}(a) and 2(b) show the spectra of FC-OSSB in 1560 and 780~nm respectively. The FC-OSSB spectra shown are acquired with a power ratio \(\omega_{s1} / \omega_{s0}\), at -3~dB at 780~nm, as desired for compensating light shifts. The \(\omega_{s-1}\) is not visible and the \(\omega_{s2}\) is approximately 21~dB below \(\omega_{s0}\). Meanwhile, concerning the 1560~nm components, the \(\omega_{-1}\) is suppressed better than 20~dB compared with \(\omega_{0}\). In the measurement, the RF power applied to the I/Q ports of the modulator is 13~dBm which corresponds to \(\beta \approx 0.74\) based on the \(V_{\pi} = 6\ \mathrm{V}\). According to the simulation in Fig.~\ref{fig:SidebandsSupress}, the sideband to carrier ratio $\omega_{s2}/\omega_{s0}$ should be below -50~dB. The degeneration can be explained by the unbalanced RF power applied on the I and Q ports and the non-identical waveguides on the two arms of the modulator.

The I/Q modulator is also used to realize suppressed carrier OSSB (SC-OSSB). Figure~\ref{fig:spctrum_OSSB}(c) shows the spectrum of SC-OSSB in 1560~nm. The \(\omega_{0}\) is suppressed to the noise level. Nevertheless the extinction ratio of \(\omega_{-1}\) is only better than -15~dB with respect to $\omega_{1}$ and the suppression of \(\omega_{s0}\) at 780~nm is considerably impaired. This is because the SFG occurs between \(\omega_{1}\) and \(\omega_{-1}\) revives \(\omega_{s0}\). The revival of the carrier at 780~nm is clearly seen in Fig.~\ref{fig:spctrum_OSSB}(d). The \(\omega_{s0}\) is only -16~dB compared with the frequency at \(\omega_{s2}\). Meanwhile, the \(\omega_{s1}\) amplitude is beneath the noise level.

\begin{figure}[htp]
	\centering
	\subfloat[]{\includegraphics[width=0.45\textwidth]{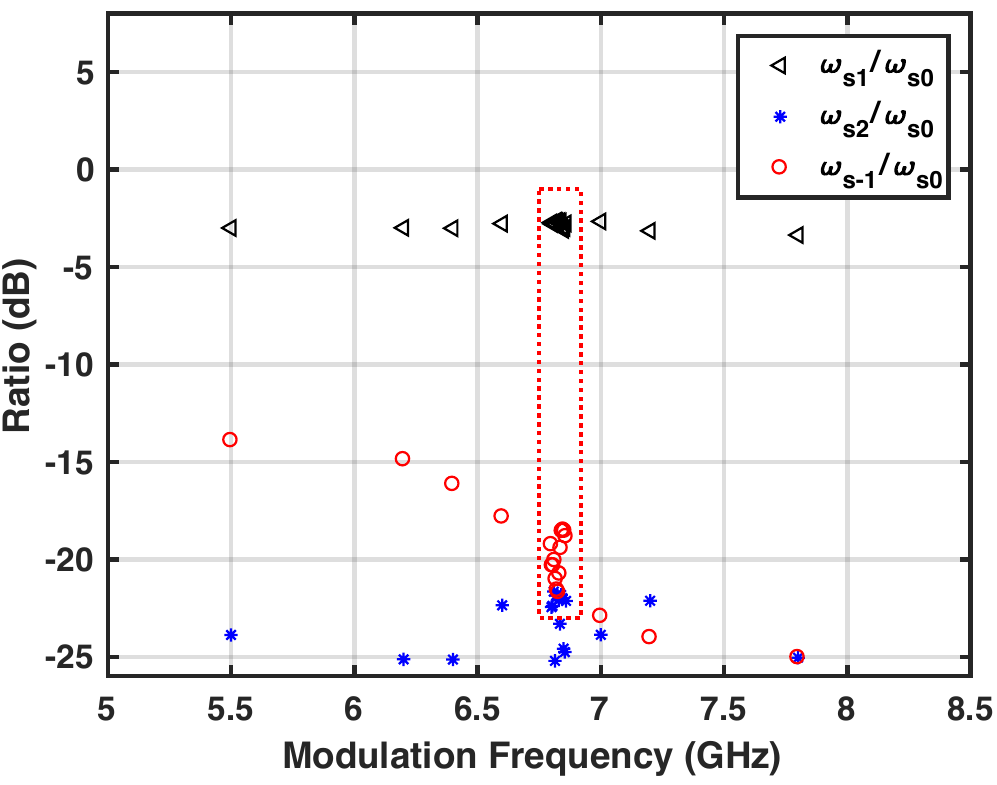}}
	\subfloat[]{\includegraphics[width=0.45\textwidth]{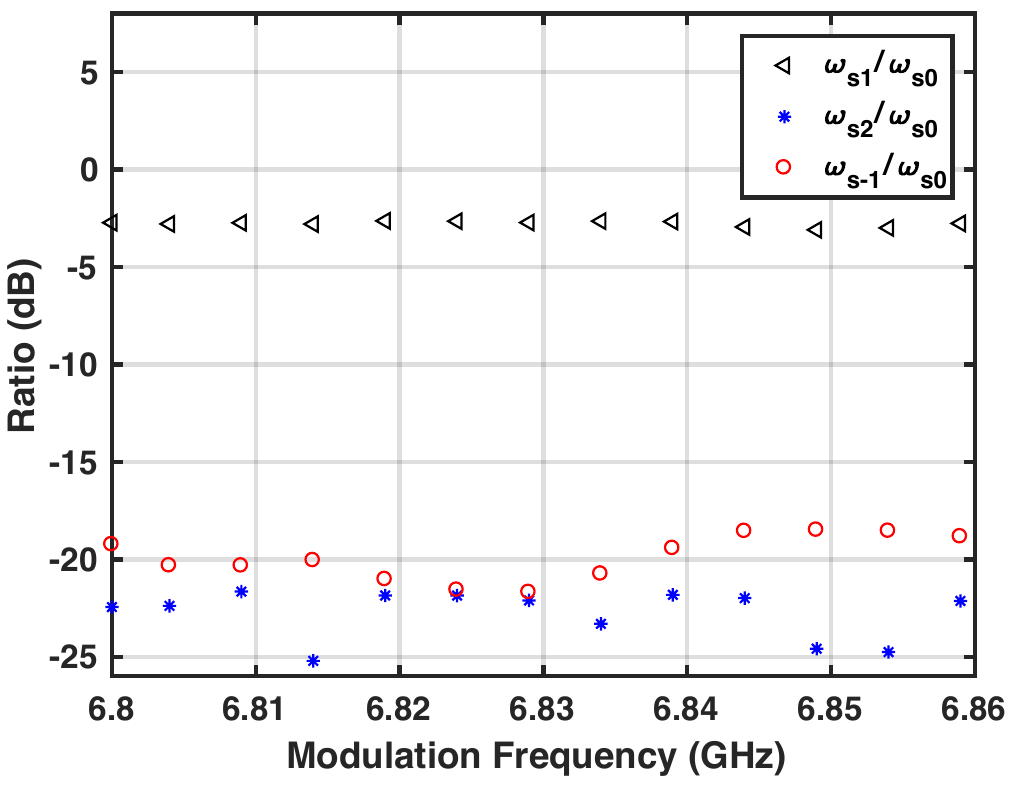}}
	\caption{\label{fig:spectrum_OSSB_780_mod_freq} The 780~nm sideband ratio verse the modulation frequency. The ratio \(\omega_{s1}/\omega_{s0}\) is to -3~dB at 6.834~GHz. The subfigure (b) is the red dash boxed region in (a).}
\end{figure}

In many applications the modulation frequency \(\omega_{m}\) needs to be tuned, for example, Doppler compensation during free fall in RAI. The tuning of \(\omega_{m}\) may produce unwelcome side effects such as degradation of suppression and power ratio variation of \(\omega_{s0} / \omega_{s1}\). These would primarily be caused by the frequency dependent characteristics of the microwave electronics driving the I/Q modulator. The dependence of the FC-OSSB performance with \(\omega_{m}\) was investigated, and the results for \(\omega_{m}\) within 4--8~GHz are shown in Fig.~\ref{fig:spectrum_OSSB_780_mod_freq}(a). Figure~\ref{fig:spectrum_OSSB_780_mod_freq}(b) shows the enlarged region covering the experimental frequency chirping for Doppler shift compensation. During scanning, the OPR of \(\omega_{s1}/\omega_{s0}\) is set to be -3~dB. The sideband \(\omega_{s2}\) can be suppressed below -20~dB between 5~GHz and 8~GHz while the suppression of the sideband \(\omega_{s-1}\) starts to increase below 6.8~GHz. This is because the RF devices have a frequency-dependent phase shift, which degrades the OSSB. However, for the range between 6.8~GHz and 6.86~GHz where the atom interferometer operates, the sideband \(\omega_{s-1}\) is suppressed below -18~dB.

\subsection{Interferometry fringe}

\begin{figure}[htp]
	\centering
	\includegraphics[width=0.8\linewidth]{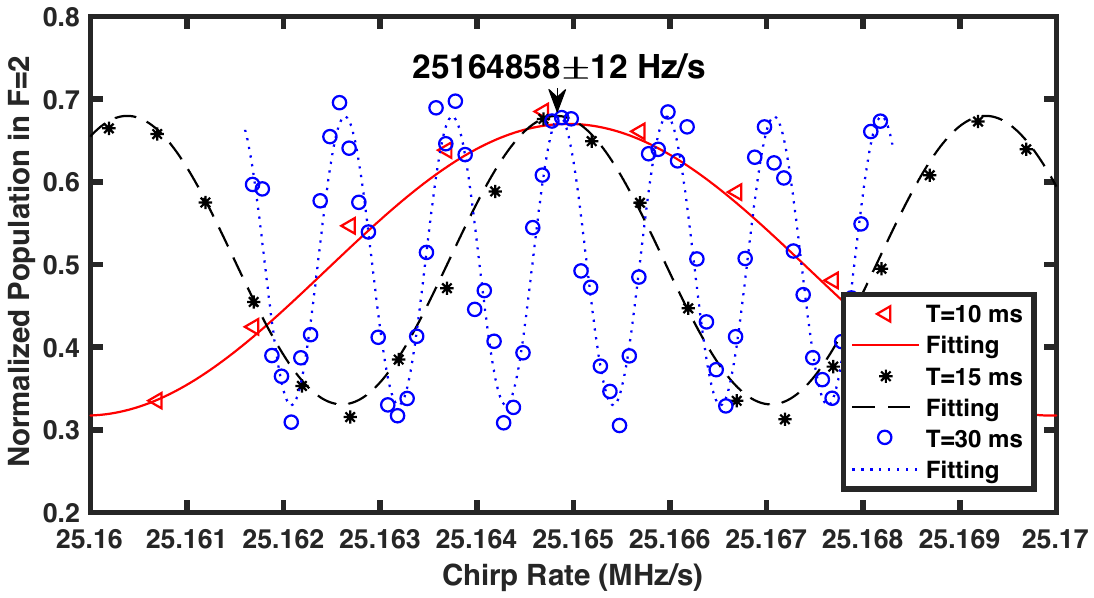}
    \caption{The interferometric fringes for determining local gravity. The chirp rate at the stationary phase point is \(25,164,858 \pm 12\,\mathrm{Hz}\), from which the local gravity can be derived to be 9.817239(4)~m/s$^{2}$.}
	\label{fig:GravityFringeChirp}
\end{figure}

The FC-OSSB laser system was then used to perform a gravity measurement in a Mach-Zehnder atom interferometer, which combines three velocity sensitive Raman pulses \cite{Peters2001}. In order to compensate the Doppler shift between the counter-propagating beams, induced by atoms in free fall, \(\omega_{m}\) is swept at a chirp rate \(\alpha\). From Eq.~(\ref{eq:phase shift}), at a specific chirp rate, the phase shift induced by the gravitational acceleration is canceled and there exists a stationary phase point independent of interferometric time \(T\). The value of \(g\) is therefore derived from the frequency chirp rate and is given by \(g = 2\,\pi\,\alpha / k_{\mathrm{eff}}\). Figure~\ref{fig:GravityFringeChirp} shows fringes with \(T\) equal to 10~ms, 15~ms and 30~ms respectively. A central fringe is addressed where the interferometer phase is canceled for a Doppler compensation. The local gravity \(g\) is determined as \(\mathrm{9.817239(4)\ m/s^{2}}\) is obtained. 

\subsection{Spatial interference elimination}
\begin{figure}[htp]
	\centering
	\includegraphics[width=0.8\linewidth]{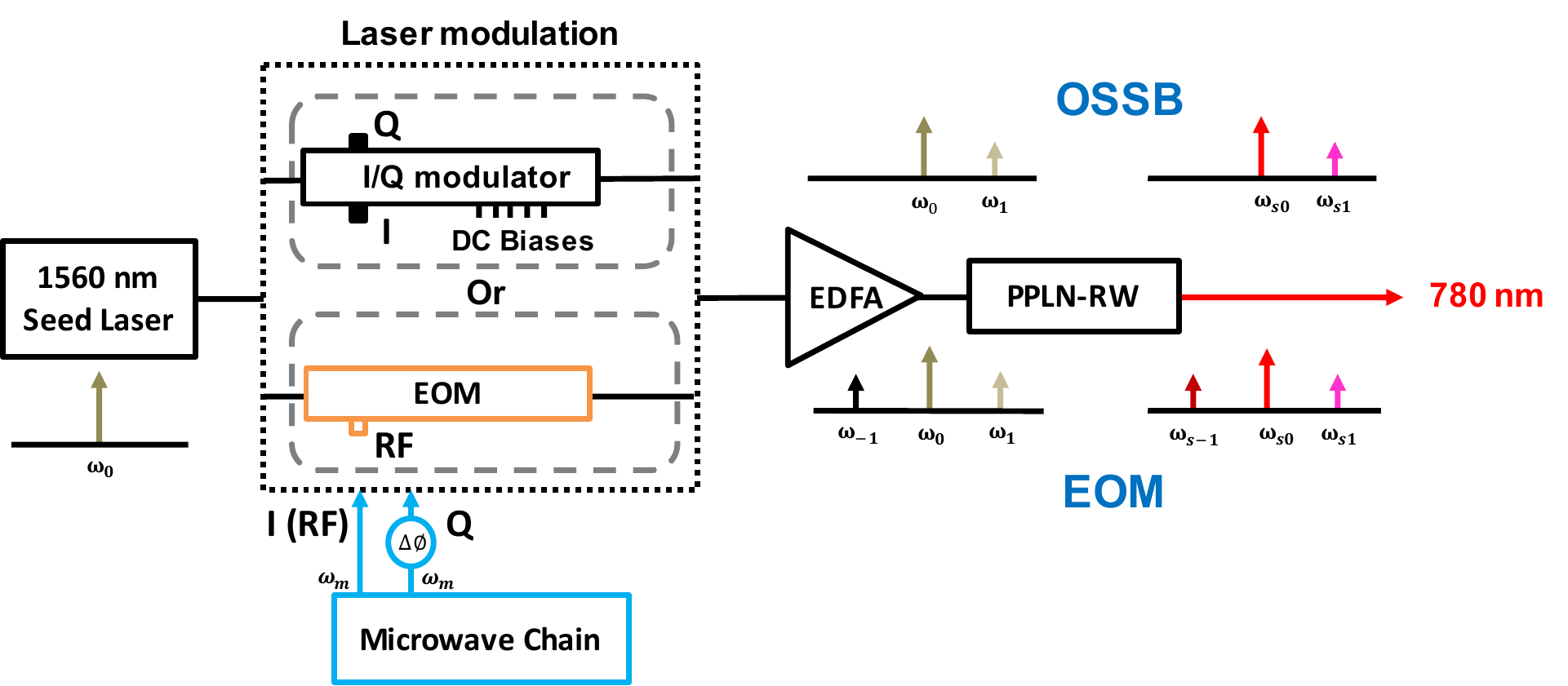}
	\caption{\label{fig:LaserScheme_SpatialEffect}The Raman laser system for measuring the spatial dependence of the Raman transition probability.}
\end{figure}

As demonstrated above, the Raman laser based on OSSB allows the suppression of the unwanted sidebands. Consequently, the interference caused by these unwanted laser lines can be eliminated. To verify this conclusion, the OSSB system is compared to an EOM scheme when applied in atom interferometry. The Raman laser setup is shown in Fig.~\ref{fig:LaserScheme_SpatialEffect}, with only the modulator changing between comparisons. 

\subsubsection{Spatially dependent Raman transition}
In EOM modulation, there are multiple pairs of frequencies that can drive resonant two-photon Raman transitions. The effective Rabi frequency contains a spatial dependence with a periodicity of \(\lambda_{\mathrm{rf}}/2\), where \(\lambda_{\mathrm{rf}}\) is the wavelength of the RF signal applied on the EOM. This causes the $\pi$ transition condition to be modified for Raman transitions performed along the interferometry region. The effect is greater when the detuning $\Delta$ is larger as the relative contribution from the unwanted frequency pairs increases~\cite{Takase2008}.

\begin{figure}[htp]
	\centering
	\includegraphics[width=0.7\linewidth]{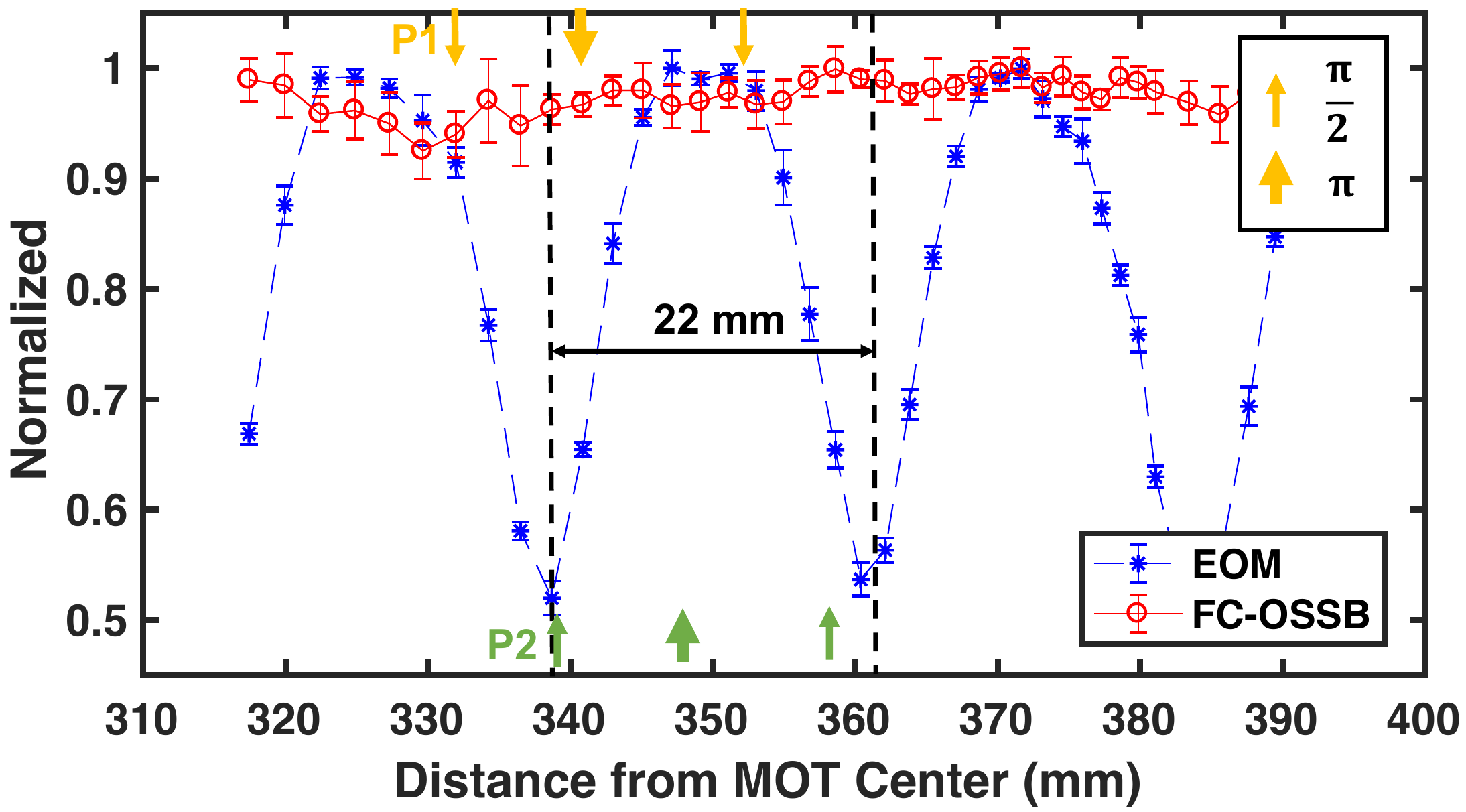}
	\caption{\label{fig:SpatialRamanTransitionCompare} The spatial dependence of the Raman transition probability. The Raman transition probabilities are measured along the beam axis with FC-OSSB and EOM Raman lasers. P1 (yellow) and P2 (green) represent two atom interferometers operate at different locations.}
\end{figure}

Raman pulses were applied at different positions along the longitudinal direction in the interferometry region. The pulse duration is set at 50~\(\mathrm{\mu s}\), which corresponds to the \(\pi\) pulse duration at the start point. The RF frequency applied to the EOM and I/Q modulator is equal to the separation between the ground state hyperfine levels F~=~1 and F~=~2 (\(\approx 6.834\ \mathrm{GHz}\)). Figure~\ref{fig:SpatialRamanTransitionCompare} shows the spatial dependence of the Raman transition. The wavelength of oscillations is measured to be 22~mm which matches with the half wavelength of the RF signal. The amplitude of the Raman transition was reduced a factor of two at the valley compared with the crest. The same measurement was repeated in the FC-OSSB scheme, showing that the spatial dependence is removed to below the experimental noise level. The fluctuation of the Raman transition is less than 10\%, which is induced by other perturbation, and shows no oscillatory spatial structure. Although the interference effect in EOM scheme can be reduced with small detuning (\(\Delta < 1\ \mathrm{GHz}\)), this increases contributions from spontaneous emission which can lead to loss the fringe contrast in atom interferometry experiments.

\subsubsection{Phase shift and contrast}
The unwanted sidebands in the EOM scheme can also induce a position dependent phase shift in atom interferometry~\cite{Carraz2012}. To evaluate the performance of the OSSB system, both schemes were used to operate a Mach-Zehnder atom interferometer at two different sets of positions, P1 and P2, which are labelled in Fig.~\ref{fig:SpatialRamanTransitionCompare}. The time separation between pulses was set to 10~ms. The sideband power ratio \(\omega_{s1} / \omega_{s0}\) is chosen to be 1/2 to cancel the first order AC Stark shift.

\begin{figure}[htp]
	\centering
	\subfloat[]{\includegraphics[width=0.7\textwidth]{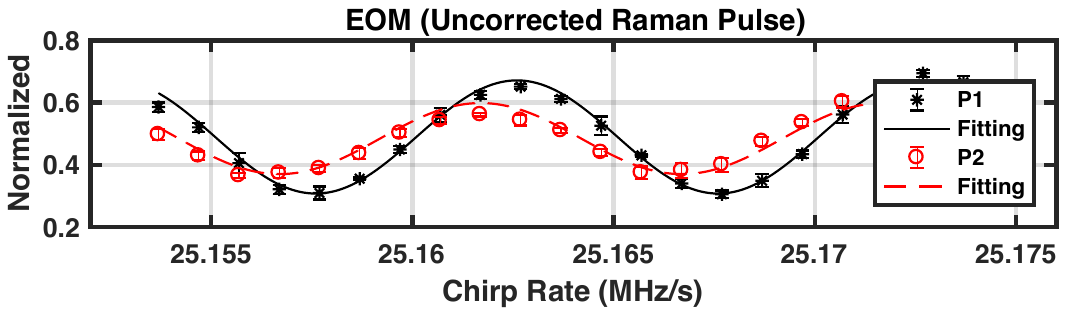}}
	\\
	\subfloat[]{\includegraphics[width=0.7\textwidth]{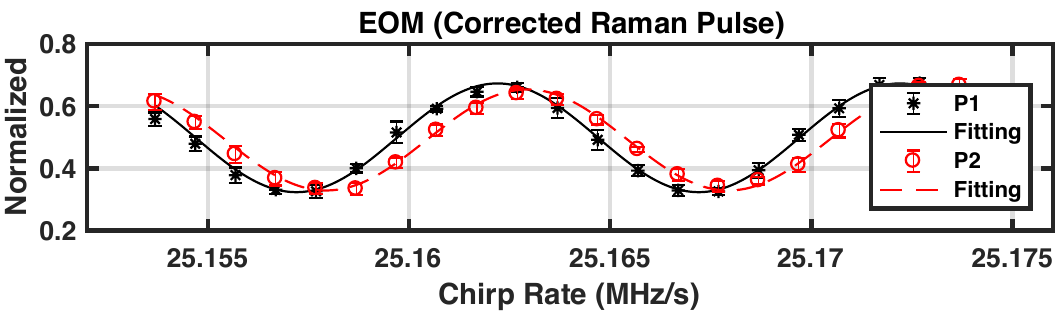}}
	\\
	\subfloat[]{\includegraphics[width=0.7\textwidth]{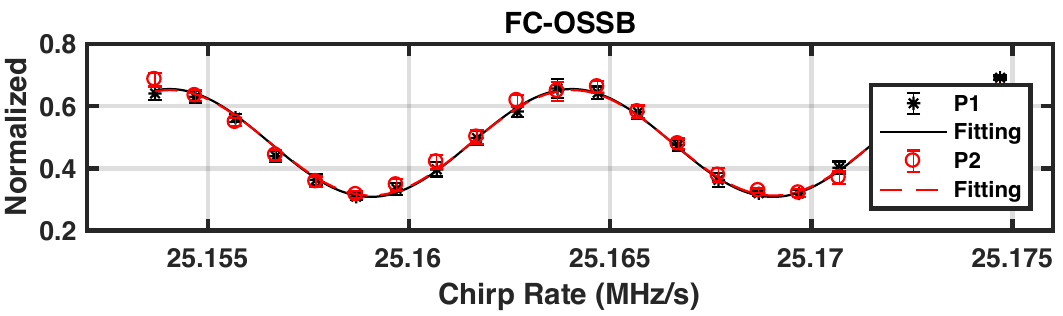}}
\caption{\label{fig:fringes_different_schemes} The interferometric fringes from different Raman laser schemes. From top to bottom, the EOM(U) scheme assuming single global Rabi frequency, the EOM(C) scheme compensating local Rabi frequency variation, and FC-OSSB.}
\end{figure}

Figure~\ref{fig:fringes_different_schemes} shows the atom interferometric fringes acquired by sweeping the chirp rate \(\alpha\) under different modulation configurations. In order to further investigate the effect arising from spatially varying Raman transition in EOM scheme, the measurements are performed under two different assumptions: (1) single global Rabi frequency; (2) spatially dependent Rabi frequency. In the first case, the Raman \(\pi\) and \(\pi/2\) pulse durations are set based on the Rabi frequency at the first position of P1. In the second case, all the pulse durations are set by the local Rabi frequency measurements. In the FC-OSSB scheme, the pulse durations are set by the Rabi frequency at the first position of P1. By fitting the data, the phase shifts and the contrasts for different conditions are extracted for comparison. The results are summarized in the table~\ref{tab:PhaseShiftandContrast}. 

\begin{table}[ht]
	\caption{\label{tab:PhaseShiftandContrast} The phase shifts and the contrasts of the interferometric fringes from different Raman laser schemes. The results for P1 and P2 are marked by black and red respectively.}
	\centering
	\begin{tabular}{c c c c c}
		\hline \hline
		 & \multicolumn{2}{c}{Phase Shift (mrad)} & \multicolumn{2}{c}{Contrast} \\
		 & P1 & P2 & P1 & P2 \\
		\hline
		EOM(U) & 1509 & \textcolor{red}{2077} & 22\% & \textcolor{red}{12\%} \\
		EOM(C) & 1759 & \textcolor{red}{1294} & 20\% & \textcolor{red}{19\%} \\
		OSSB & 597 & \textcolor{red}{596} & 21\% & \textcolor{red}{20\%} \\
\hline \hline
\end{tabular}
\end{table}

With the EOM scheme, the fringes measured at P1 and P2 shift away from the fringes measured with the FC-OSSB scheme, regardless of whether the Raman pulses duration is corrected or not. Without correcting the Raman pulse duration in the EOM scheme, the phase measured at P1 and P2 has a relative phase difference 568~mrad. In addition, the contrast is reduced from 22\% to 12\%, nearly a factor of two decrease between P1 and P2. After correcting the Raman pulse duration, the fringe contrast at P2 can be recovered to 19\% but there is still a spatially dependent phase difference at both P1 and P2. However, with the FC-OSSB scheme, the fringe at P2 is shifted by  1~mrad with a small contrast decrease around 1\% compared with the one at P1. As proposed in the paper~\cite{Carraz2012}, the EOM scheme is still adoptable for atom interferometry application in high-precise gravity measurement by numerically calculate the phase shift induced by the unwanted laser lines in the EOM scheme, which demands the atom interferometer parameters are precisely estimated. However, the OSSB approach has been shown to effectively remove these concerns, and others, without adding additional complexity to the laser system.

\section{Conclusion}
The application of FC-OSSB in atomic physics, especially RAI, is demonstrated in this paper. The suppression of the spatially dependent Rabi frequency and the interferometric phase bias induced by undesired sidebands, which arise through existing single laser schemes, are shown to be suppressed to a level below the experimental noise of the atom interferometer. While enabling these improvements, the compact, robust and single laser nature of the EOM scheme is retained - allowing for the improvement of compact RAI based sensing. In addition, carrier suppression (SC-OSSB) to create an agile single frequency is also demonstrated, enabling future implementation as an efficient and broadband optical frequency shifter with the potential to significantly improve over techniques such as serrodyne modulation~\cite{Johnson2010}. By aid of versatile microwave electronics, in principle, OSSB schemes can be combined to realize a broadly tunable single-laser light source for RAI.

\section*{Appendix: I/Q modulator based OSSB}\label{sec:Appendix}

\begin{figure}[htp]
	\centering
 	\includegraphics[width=0.7\linewidth]{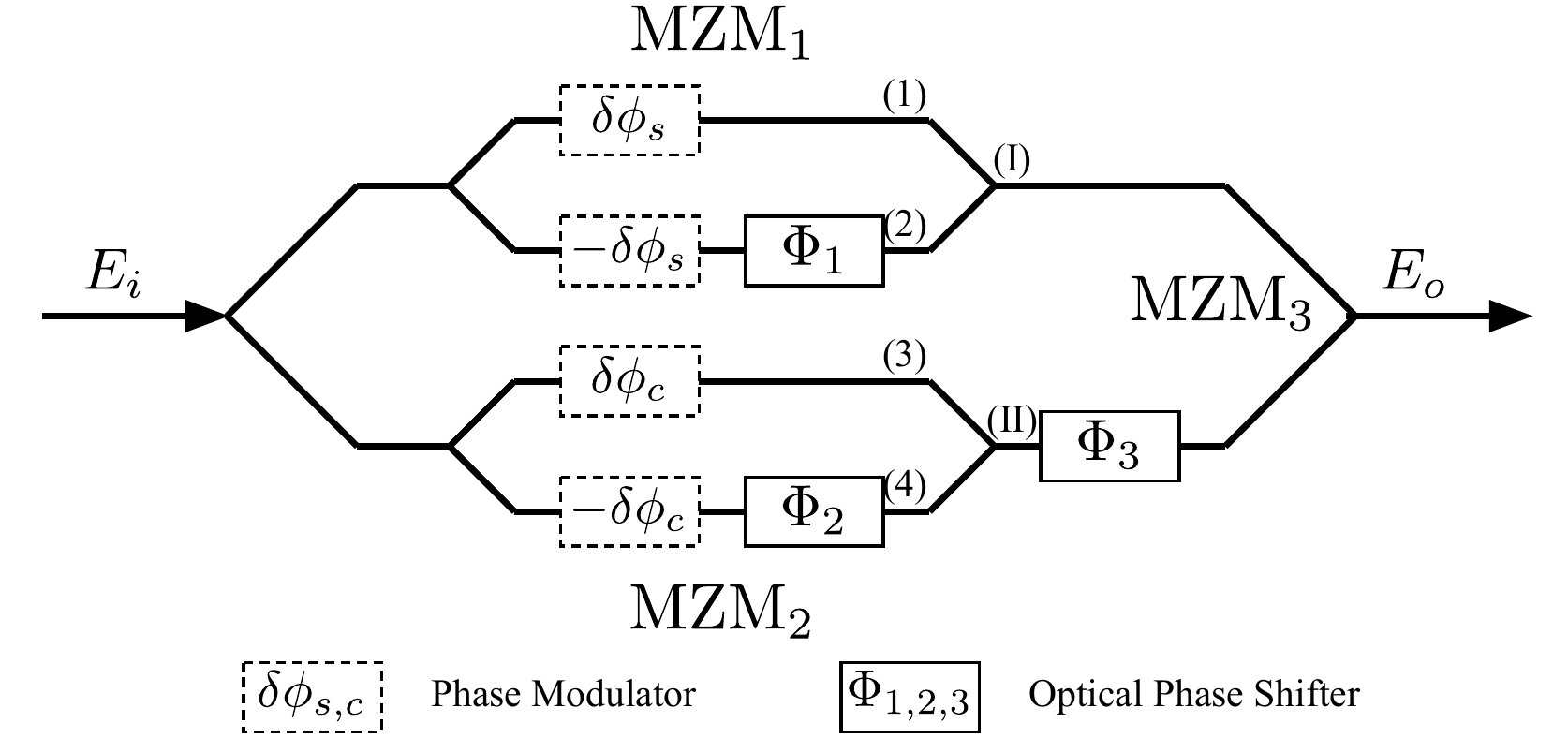}
    \caption{\label{fig:IQStructure} Illustration of the simplified structure of a dual parallel Mach-Zehnder modulator. Sine phase modulator: \(\Delta \phi_{s} = \beta \sin{\omega_m t}\). Cosine phase modulator: \(\Delta \phi_{c} = \beta \cos{\omega_m t}\). \(\mathrm{\Phi_{1, \, 2, \, 3}}\): optical phase shifter.}
\end{figure}

Our I/Q modulator is based on a dual parallel Mach-Zehnder modulator (MZM)~\cite{Izutsu1981, Shimotsu2001}. The simplified structure is shown in Fig.~\ref{fig:IQStructure}. Sub-modulators, \(\mathrm{MZM}_{1}\) and \(\mathrm{MZM}_{2}\), act as either single or double balance modulators controlled by the phase shifters~\(\Phi_{1,2}\). A third sub-modulator \(\mathrm{MZM}_{3}\) is essentially a combiner for \(\mathrm{MZM}_{1,2}\) with an extra phase shifter~\(\Phi_{3}\). Each arm of \(\mathrm{MZM}_{1, 2}\) is simply a conventional phase modulator. Assuming the seed laser frequency~\(\omega_{c}\) and a modulation signal \(\omega_{m}\) with modulation depth~\(\beta\), the signals at locations (1)--(4) can be expressed by Bessel functions:
\begin{equation}\label{eq:PM}
\begin{split}
    E_{o} & = E_{(1)} + E_{(2)} + (E_{(3)} + E_{(4)})\,e^{i\,\Phi_{\mathrm{3}}} \\
          & = \frac{E_{i}}{2}\,e^{i\,\omega_{c} t}\,[e^{i(\beta\,\sin{\omega_{m} t})} + e^{i(-\beta\,\sin(\omega_{m} t) + \Phi_{1})} + e^{i(\beta\,\cos{(\omega_{m} t)} + \Phi_{3})} + e^{i(-\beta\,\cos{(\omega_{m} t)} + \Phi_{2} + \Phi_{3})}]\\
          & = \frac{E_{i}}{2}\,e^{i\,\omega_{c} t}\,\sum_{-\infty}^{\infty}\,[J_{n}(\beta)\,e^{i\,n\,\omega_{m} t} + J_{n}(-\beta)\,e^{i(\,n\,\omega_{m} t + \Phi_{1})} + J_{n}(\beta)\,e^{i\,(n\,(\omega_{m} t + \pi/2) + \Phi_{3})} \\
          & + J_{n}(-\beta)\,e^{i\,(n\,(\omega_{m} t + \pi/2) + \Phi_{2} +\Phi_{3})}] \\
          & = \frac{E_{i}}{2}\,e^{i\omega_{c}t}\,\{C_{0}\,J_{0}(\beta) + \sum_{n=1}^{\infty}\,[A_{n}\,J_{n}(\beta)\,e^{i\,n\,\omega_{m} t} + B_{n}\,J_{n}(\beta)\,e^{-i\,n\,\omega_{m} t}]\}\\
\end{split}
\end{equation}
where $J_{n}(\beta)$ denotes the $n$th-order Bessel function and the coefficients $C_{0}$, $A_{n}$ and $B_{n}$ are expressed as below: 
\begin{eqnarray}
    C_{0} & = & 1 + e^{i\Phi_{1}} + e^{i\Phi_{3}}\,(1 + e^{i\Phi_{2}}) \\
    A_{n} & = & 1 + (-1)^{n}\,e^{i\Phi_{1}} + e^{i \Phi_{3}}\,[e^{i \frac{n \pi}{2}} + (-1)^{n}\,e^{i(\frac{n \pi}{2} + \Phi_{2})}] \\
    B_{n} & = & (-1)^{n} + e^{i\Phi_{1}} + e^{i \Phi_{3}}\,[(-1)^{n}\,e^{-i \frac{n \pi}{2}} + e^{i(-\frac{n \pi}{2} + \Phi_{2})}]
\end{eqnarray}

The FC-OSSB can be achieved by setting \(\Phi_{1,2,3} = \pi/2, \pi/2, 3\pi/2\):

\begin{equation}\label{eq:FC-OSSB}
\begin{split}
    E_{o} =\frac{E_{i}}{2}[ &\ 2\,J_{0}(\beta)\,e^{i \omega_{c} t} \\
            & + 2\,(1 - e^{i \pi/2})\,J_{1}(\beta)\,e^{i (\omega_{c} + \omega_{m}) t} \\
            & + 2\,e^{i \pi/2}\,J_{2}(\beta)\,e^{i (\omega_{c} + 2 \omega_{m}) t} + 2\,e^{i \pi/2}\,J_{1}(\beta)\,e^{i (\omega_{c} - 2\omega_{m}) t} \\
            & - 2\,(1 - e^{i \pi/2})\,J_{3}(\beta)\,e^{i (\omega_{c} - 3 \omega_{m}) t} \\
            & + 2\,J_{4}(\beta)\,e^{i (\omega_{c} + 4 \omega_{m}) t} + 2\,J_{4}(\beta)\,e^{i (\omega_{c} - 4 \omega_{m}) t} + \ldots]
\end{split}
\end{equation}

We also can realize the suppressed-carrier optical single sideband (SC-OSSB) by setting \(\Phi_{1,2,3}\) to \(\pi, \pi, 3\pi/2\):
\begin{equation}\label{eq:SC-OSSB}
    E_{o} = \frac{E_{i}}{2}[4\,J_{1}(\beta)\,e^{i (\omega_{c} + \omega_{m}) t} - 4\,J_{3}(\beta)\,e^{i (\omega_{c} - 3\omega_{m}) t} + 4\,J_{5}(\beta)\,e^{i (\omega_{c} + 5\omega_{m}) t} + \ldots]
\end{equation}

\subsection*{OSSB with second harmonic generation}
The spectral coverage of OSSB can be further extended to where there are no convenient light sources or modulators available by nonlinear optical processes such as SHG. Unlike other nonlinear processes, the SHG of OSSB will unavoidably degrade the OSSB because of the self-mixing characteristics. The self-mixing process pairs the different frequency components of OSSB and creates undesired frequencies. For simplicity, we assume \(n = 4\). In the case of FC-OSSB, the SHG signal can be expressed as:

\begin{equation}\label{equ:FC-OSSB-780}
\begin{split}
^{FC}E_{S} \propto & \epsilon_{0} \chi^{2} E_{o}^{2}\\
\approx&\dfrac{\epsilon_{0} \chi^{2}E_{i}^{2}}{4}\lbrace[(4J_{0}^{2}(\beta)-8J^{2}_{2}(\beta))]e^{i\omega_{s0}t}\\
&+8(1-e^{i\pi/2})J_{0}(\beta)J_{1}(\beta)e^{i\omega_{s1}t}+8(1+e^{i\pi/2})J_{2}(\beta)J_{1}(\beta)e^{i\omega_{s-1}t}\\
&+8e^{i\pi/2}[J_{0}(\beta)J_{2}(\beta)-J_{1}(\beta)J_{1}(\beta)]e^{i\omega_{s2}t}+8e^{i\pi/2}[J_{0}(\beta)J_{2}(\beta)-2J_{3}(\beta)J_{1}(\beta)]e^{i\omega_{s-2}t}\\
&+8(1+e^{i\pi/2})J_{1}(\beta)J_{2}(\beta)e^{i\omega_{s3}t}+8(1-e^{i\pi/2})J_{0}(\beta)J_{3}(\beta)e^{i\omega_{s-3}t}\\
&+[8J_{0}(\beta)J_{4}(\beta)-4J^{2}_{2}(\beta)]e^{i\omega_{s4}t}+[8J_{0}(\beta)J_{4}(\beta)-4J^{2}_{2}(\beta)]e^{i\omega_{s-4}t}\rbrace\\
\end{split}
\end{equation}
where $\epsilon_{0}$ is the permittivity of vacuum and $\chi$ is the susceptibilities of the medium. The shorthand for the different frequency components is as:
\begin{equation*}
	\begin{split}
	\omega_{n} & = \omega_{c} + n\,\omega_{m} \\
	\omega_{0} & = \omega_{c},\ \omega_{1} = \omega_{c} + \omega_{m},\ \omega_{-1} = \omega_{c} - \omega_{m} \ldots
	\end{split}
\end{equation*}
and "s" is added to the subscript to refer the components after SHG:
\begin{equation*}
	\begin{split}
	\omega_{sn} & = 2 \omega_{c} + n\,\omega_{m} \\
	\omega_{s0} & = 2 \omega_{c},\ \omega_{s1} = 2 \omega_{c} + \omega_{m},\ \omega_{s-1} = 2 \omega_{c} - \omega_{m} \ldots
	\end{split}
\end{equation*}
Figure~\ref{fig:SidebandsSupress} shows the simulation of the optical power ratio (OPR) of the sidebands with respect to the carriers in 780 nm versus the modulation depth $\beta$. Assuming \(\beta = 0.58\), the power ratio of \(\omega_{s1}$/$\omega_{s0}\) is -3~dB while the other sidebands are suppressed better than -40~dB.

Similarly, the SHG of SC-OSSB can be expressed as:
\begin{equation}
\begin{split}
^{SC}E_{S} \propto & \epsilon_{0} \chi^{2} E_{o}^{2}\\
=&\dfrac{\epsilon_{0} \chi^{2}E_{i}^{2}}{4}[16J_{1}(\beta)J_{1}(\beta)e^{i\omega_{s2}t}+32J_{3}(\beta)J_{1}(\beta)e^{i\omega_{s-2}t}]
\end{split}
\end{equation}

It is independent of the modulation depth, meaning that we do not need to over-drive the modulator to achieve carrier suppression.

\section*{Funding}
\href{http://dx.doi.org/10.13039/501100000266}{Engineering and Physical Sciences Research Council} (EPSRC) (EP/M013294); \href{http://dx.doi.org/10.13039/100010418}{Defence Science and Technology Laboratory} (Dstl) (DSTLX-1000095040); \href{http://dx.doi.org/10.13039/100010664}{Future and Emerging Technologies} within the 7th Framework Programme for Research of the European Commission (FET) (FP7-ICT-601180)

\end{document}